\documentclass[11pt]{article}
\pdfoutput=1

\usepackage{jheppub}
\usepackage{rotating}
\usepackage{enumitem}
\usepackage{wrapfig}
\usepackage[left=5cm,right=0cm,top=5cm,bottom=1cm,footskip=1cm]{geometry}

\usepackage{amsfonts}
\usepackage{amsmath}
\usepackage{slashed}
\usepackage{amssymb}
\usepackage{latexsym}
\usepackage{multirow}
\usepackage{hyperref}
\usepackage{enumitem}
\usepackage{mathtools}
\hypersetup{colorlinks=true,citecolor=red,linkcolor=magenta,urlcolor=blue}
\usepackage{relsize}
\usepackage{booktabs}
\usepackage{subfigure}
\usepackage{cleveref}
\usepackage{fancyhdr}
\usepackage{float}
\usepackage{graphicx}
\usepackage{microtype}
\usepackage{tabularx}
\usepackage{xcolor}
\usepackage{orcidlink}
\usepackage{wrapfig}
\usepackage{siunitx,adjustbox,bm}
\usepackage[bottom]{footmisc}
\usepackage{physics}
\newcolumntype{C}{>{$}c<{$}}
\newcolumntype{L}{>{$}l<{$}}
\newcolumntype{R}{>{$}r<{$}}

\title{Phenomenology of a Kinetic Higgs Portal}
\author[a,b]{Anisha\orcidlink{0000-0002-5294-3786},}
\author[c]{Lisa Biermann\orcidlink{0000-0003-1469-6400},} 
\author[d]{Christoph Englert\orcidlink{0000-0003-2201-0667},} 
\author[a]{and Margarete M\"uhlleitner\orcidlink{0000-0003-3922-0281}}
\affiliation[a]{Institute for Theoretical Physics, Karlsruhe Institute of Technology, Wolfgang-Gaede-Str. 1, 76131 Karlsruhe, Germany}
\affiliation[b]{Institute for Astroparticle Physics, Karlsruhe Institute of Technology, 76344 Eggenstein-Leopoldshafen, Germany}
\affiliation[c]{PSI Center for Neutron and Muon Sciences, 5232 Villigen PSI, Switzerland}
\affiliation[d]{Department of Physics \& Astronomy, University of Manchester, Oxford Road, Manchester M13 9PL, United Kingdom}

\emailAdd{anisha@kit.edu}
\emailAdd{lisa.biermann@psi.ch}
\emailAdd{christoph.englert@manchester.ac.uk}
\emailAdd{margarete.muehlleitner@kit.edu}
\preprint{KA-TP-30-2025}
\abstract{We explore the phenomenological consequences of non-minimal hidden sector interactions on observable correlations in the Higgs sector, mediated through the $\mathbb{Z}_2$-symmetric Higgs portal. Particular attention is given to non-standard momentum dependencies of the hidden sector scalar, which arise naturally in an effective field theory (EFT) framework, e.g. in Composite Scalar Dark Matter theories. We discuss the implications of such hidden sector interactions for the thermal history of the universe. We show that aspects of such non-standard momentum dependencies can be probed at future lepton colliders such as a FCC-ee, potentially also through radiative corrections. This gives rise to precision probes for regions where direct detection constraints and relic abundance can be accounted for as predicted in, e.g., Composite Scalar Dark Matter theories.}
\begin{document}
%%%%%%%%%%%%%%%%%%%
\maketitle
\flushbottom
\allowdisplaybreaks
%%%%%%%%%%%%%%%%%%%
\section{Introduction}
\label{sec:intro}
%%%%%%%%%%%%%%%%%%%
Searches for physics beyond the Standard Model (BSM) at CERN's Large Hadron Collider (LHC), albeit unsuccessful so far, give crucial information about the microscopic origin of the electroweak scale. It is widely established that BSM phenomena are required to address critical shortfalls of the Standard Model (SM), whether this concerns the theoretical self-consistency of the SM (through fine-tuning) or agreement with experimental results (e.g. related to dark matter and baryogenesis). On the one hand, in generic approaches to uncover sensitivity to new physics, deviations from the SM are parametrised using effective field theory (EFT), either in the linear (SMEFT)~\cite{Grzadkowski:2010es} or non-linear (HEFT) versions, cf.~e.g.,~\cite{Longhitano:1980tm,Feruglio:1992wf,Appelquist:1993ka,Alonso:2012px}. On the other hand, specific and popular extensions of the SM are so-called Higgs portal interactions that seek to marry the dark sector phenomena with the electroweak scale in the SM~\cite{Binoth:1996au,Patt:2006fw,Schabinger:2005ei}. These leverage one of three renormalisable options to connect the SM to a hidden sector (in addition to kinetic $U(1)$ mixing and mixing with a right-handed sterile neutrino).

The relevance of Higgs portal interactions for the LHC and beyond has been widely researched in the literature (e.g.~\cite{Barger:2007im,He:2009yd,Englert:2011yb}); their appeal for driving a strong first-order phase transition is well-understood~\cite{Profumo:2007wc,Biermann:2022meg}. Furthermore, the ${\mathbb{Z}}_2$-symmetric portal model enables a transparent comparison of present and future collider performances across the precision/energy coverage domains that we can expect at such machines (see in particular~\cite{Ruhdorfer:2019utl,Englert:2020gcp}). 

It is conceivable that a hidden sector exhibits a richer phenomenology~\cite{Gross:2017dan,Criado:2021trs} than parametrised by the minimal Higgs portal. Unless there are sizable non-linearities on the visible side~\cite{Brivio:2015kia}, when the hidden sector can be integrated out, it will by construction impart SMEFT-like patterns, even when the hidden sector is highly non-linear. We will see that these interactions are, in some sense, then maximally HEFT-like from the point of view of convergence (see also~\cite{Helset:2020yio}). Nonetheless, the phenomenology will be dominantly visible through the interactions of the Higgs boson, motivating the language of HEFT to capture these effects transparently, in particular when the hidden degrees of freedom are light~\cite{Anisha:2023ltp}. Furthermore, effective portal interactions will enlarge the model's parameter region, phenomenologically extending beyond the renormalisable correlation expectations.

In this work, we aim to provide a phenomenological exploration of momentum-dependent, effective portal interactions beyond leading order in the visible sector. These extend the traditional portal interactions into the realm of effective field theory. Such interactions have been considered more broadly in~\cite{Song:2023jqm,Song:2023lxf,Delaunay:2025lhl} from a bottom-up perspective. Conversely, such interactions have been motivated first in scenarios of Composite Scalar Dark Matter~\cite{Frigerio:2012uc}. In particular, we focus on local momentum-dependent 
interactions in the hidden sector that are communicated to the visible sector in a standard way $\sim \Phi^\dagger \Phi$ (with $\Phi$ denoting the SM Higgs doublet). Section~\ref{sec:momdep} gives a brief, qualitative motivation on how such interactions can appear in a range of strongly-interacting sectors. In Section~\ref{sec:collider}, we turn to collider probes of such interactions with a particular emphasis on how expected SM correlation patterns are affected by the virtual presence of such physics. Non-minimal interactions
%%%%%%%%%%%%%%%%%%%%%%%%%%%%%%%%%%%%%%%%%%%%%%%%%
\begin{wrapfigure}[10]{r}{0.36\textwidth}
\hfill\parbox{0.34\textwidth}{
\centering
\includegraphics[width=0.24\textwidth]{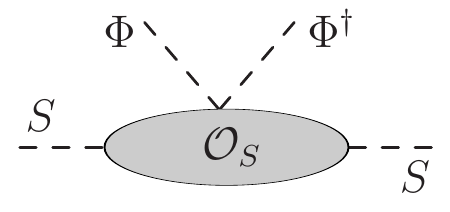}
\caption{Generic non-minimal ${\mathbb{Z}}_2$-symmetric Higgs portal interactions coupling two emergent scalar $S$ bosons to the SM Higgs doublet $\Phi$ via interactions ${\cal{O}}_S$ at the lowest multiplicity order.
\label{fig:hidden1}}}
\end{wrapfigure}
%%%%%%%%%%%%%%%%%%%%%%%%%%%%%%%%%%%%%%%%%%%%%%%%%%
can also open up a parameter region where invisible Higgs decays remain in agreement with current Higgs physics constraints. These regions can be relevant for the aforementioned Composite Scalar Dark Matter theories. Finally, in Section~\ref{sec:cosm}, we clarify how the presence of such interactions can induce or modify a first-order phase transition in the early universe before concluding in Sec.~\ref{sec:conc}.

%%%%%%%%%%%%%%%%%%%%%%%%%%%%%%%%%%%%%%%%%%%%%%%%%%
\section{Effective momentum dependencies}
\label{sec:momdep}
%%%%%%%%%%%%%%%%%%%%%%%%%%%%%%%%%%%%%%%%%%%%%%%%%%
\newcommand{\smalloverleftrightarrow}[1]{\overset{\scriptscriptstyle\leftrightarrow}{#1}}
As a straightforward example of an effective Higgs portal to strong sectors, consider, e.g., a hidden, QCD-like sector below the symmetry-breaking scale (see also~\cite{Hur:2007uz,Hur:2011sv}). Its phenomenology will be characterised by a set of pseudo-Nambu Goldstone bosons parametrised by a Callan-Coleman-Wess-Zumino field of the coset $G/H$ (taken as a symmetric space for simplicity), $\xi=\exp(i\pi/f)$~\cite{Callan:1969sn,Coleman:1969sm}. A Higgs portal, coupled to this sector, can then exhibit low-energy effective interactions (see also~\cite{Hatanaka:2016rek,Choi:2018iit}), Fig.~\ref{fig:hidden1},
\begin{equation}
\label{eq:sqcd}
{\cal{L}}^{\text{EFT}}={\cal{L}}_{\text{SM}} - {\Phi^\dagger \Phi } \left(
{f^2\over \,4  \Lambda_1^2} \text{Tr}[M(\xi+\xi^{\dagger})] +  {f^2\over \,4 \Lambda_2^2} \text{Tr}(\partial_\mu \xi^\dagger \partial^\mu \xi ) + \dots \right),
\end{equation}
where $M$ is an explicit source of symmetry violation (cf.~the light quark flavour masses in QCD) compatible with the ${\mathbb{Z}}_2$-odd symmetry assignment of the would-be pion fields except for a possible Wess-Zumino-Witten term~\cite{Wess:1971yu,Witten:1983tw} (see also~\cite{Frigerio:2012uc}). Momentum dependencies of the effective theory below the cut-off $\Lambda_2$ are therefore expected to emerge. These theories have been more broadly motivated and studied in a series of publications on Composite Scalar Dark Matter~(see, e.g., \cite{Marzocca:2014msa, Carmona:2015haa,Bruggisser:2016ixa,Bruggisser:2016nzw,Balkin:2018tma} for relevant studies). In the following, we will approach our analysis from the perspective of EFT without making specific references to potential UV completions. However, the relation to Composite Scalar Dark Matter theories becomes relevant when discussing dark matter constraints further below. 

To observe visibly large corrections due to non-standard momentum dependencies in the hidden sector, we need the latter to be the a priori dominant effects. We briefly sketch how such a situation can arise in large-$N$ theories. For a more careful treatment of models in which such interactions are sourced in parallel to a range of additional new physics effects in the Higgs sector, we refer the reader to the existing composite literature~\cite{Frigerio:2012uc,Marzocca:2014msa, Bruggisser:2016ixa,Bruggisser:2016nzw,Balkin:2018tma}.

Consider, e.g., the (unrenormalised) two-point function of the hidden-sector scalar $S$. It can be expressed through a K\"all\'en-Lehmann representation~\cite{Kallen:1952zz,Lehmann:1954xi} 
\begin{equation}
\label{eq:kl}
\left\langle 0 | T\{{{S}}(x){{S}}(0) \}  |0 \right\rangle = i \int_0^\infty \text{d} q^2 \, \rho_S(q^2) \int {\text{d}^4 p\over (2\pi)^4}  \, {e^{-ip\cdot x}\over p^2- q^2 +i\epsilon}\,,
\end{equation}
with the spectral density $\rho_S(q^2)$ of states excited from the vacuum by the quantum field $S$. The standard single particle-scenario in this parametrisation is obviously recovered through identifying $\rho_S(q^2) \to \delta(q^2-m_S^2) | \langle 0 | S(0) | q \rangle |^2$ with hidden-sector state excitations characterised by $p_S^2=m_S^2$. But less canonical representations are possible. For instance, Georgi considered `unparticle' spectral densities in~\cite{Georgi:2007ek} 
\begin{equation}
\label{eq:unparticle}
2\pi\,\theta(q^2) \theta(q^0) \,\rho_S(q^2) = A_{d_{\cal{U}}} \,\theta(q^2) \theta(q^0) \,q^{2d_{\cal{U}}-4}=
{16\pi^{5/2}\over (2\pi)^{2{d_{\cal{U}}}}} {\Gamma(d_{\cal{U}}+1/2) \over \Gamma(d_{\cal{U}}-1) \Gamma(2d_{\cal{U}})}
\theta(q^2) \theta(q^0) 
\left(q^2\right)^{d_{\cal{U}}-2},
\end{equation}
identifying the right-hand side of Eq.~\eqref{eq:unparticle} with the phase space of a non-integral number $d_{\cal{U}}$ of massless particles. This is motivated by understanding $S$ as a conformal operator with scaling dimension $d_{\cal{U}}$; such a theory naturally lends itself to AdS/CFT duality descriptions~\cite{Arkani-Hamed:2000ijo,Klebanov:1999tb,Cacciapaglia:2008ns}. Entering Eq.~\eqref{eq:unparticle} into Eq.~\eqref{eq:kl} leads to a non-local behaviour $\sim x^{-2d_{\cal{U}}}$, for $1\leq d_{\cal{U}}<2$
\begin{equation}
\label{eq:prop}
 -i\left\langle 0 | T\{{{S}}(x){{S}}(0) \}  |0 \right\rangle = {A_{d_{\cal{U}}} \over 2 \sin\left(d_{\cal{U}}\pi\right)} \int {\text{d}^4 p\over (2\pi)^4} {e^{-ip\cdot x}\over (-p^2-i\epsilon)^{2-d_{\cal{U}}}} \stackrel{d_{\cal{U}}\to 1}{\longrightarrow} \int {\text{d}^4 p\over (2\pi)^4} {e^{-ip\cdot x}\over p^2 + i\epsilon}\,,
\end{equation}
reducing to a massless propagator for unity operator dimensions.

Fox, Rajaraman, and Shirman~\cite{Fox:2007sy} proposed a modification $q^2 \to q^2-\mu^2$ of Eq.~\eqref{eq:unparticle}, introducing an infrared cut-off $\mu^2$ of the original, continuous CFT spectrum to obtain more realistic theories. This also yields the standard scalar propagator for $d_{\cal{U}}\to 1$ and potentially additional poles appearing in the spectrum. The extension to $d_{\cal{U}}>2$ is non-trivial and has been discussed in~\cite{Cacciapaglia:2008ns}: The divergent behaviour of the two-point function in this instance requires local subtraction terms, which can quickly dominate the phenomenology as the unparticle spectrum decouples (see~also~\cite{Cacciapaglia:2007jq}); such theories are sensitive to the UV cut-off via the subtraction terms.

This discussion can be qualitatively extended to $2\to 2$ scattering amplitudes that also obey dispersion relations. A generic Higgs portal interaction
\begin{equation}
\label{eq:hidd}
{\cal{L}}={\cal{L}}_{\text{SM}} - \eta\, \Phi^\dagger \Phi \,{\cal{O}}^2_S\,,
\end{equation}
with a ${\mathbb{Z}}_2$-odd SM-singlet and scalar operator ${\cal{O}}_S$ can exhibit a highly non-trivial interaction and momentum dependencies. In cases of large anomalous operator dimensions, the above holographic interpretations can be employed to gain qualitative insights. In this case, the three-point function is fixed by conformal symmetry~\cite{Witten:1998qj}. Defining the portal ${\cal{O}}_{\text{SM}}\sim \Phi^\dagger \Phi$, $x_{ij}=x_i-x_j$, and taking the {\emph{simultaneous}} limit $|x_{ij}|=z\to 0$, 
\begin{equation}
\langle 0 | {\cal{O}}_{\text{SM}}(x_1) {\cal{O}}_{S}(x_2) {\cal{O}}_{S}(x_3)| 0 \rangle
\sim {1\over |{x_{12}}|^2 |{x_{13}}|^2 |{x_{23}}|^{-2+2d_{\cal{U}}}} 
\longrightarrow {1\over z^{2+2d_{\cal{U}}}}\,,
\end{equation}
expectedly displaying a similar non-local behaviour as in Eq.~\eqref{eq:prop}. (This depends on the path chosen to apply the operator product expansion.) With subtractions (e.g. to satisfy the Froissart bound~\cite{Froissart:1961ux}), again the interactions are dominated by local terms of an interpolating field $S$ 
\begin{subequations}
\label{eq:heftex}
\begin{equation}
{\cal{L}}^{\text{EFT}}={\cal{L}}_{\text{SM}} - \Phi^\dagger \Phi \left(
{\eta_S\over 2} S^2 + {\eta_{KS}\over \Lambda^2} \partial_\mu S \partial^\mu S + \dots \right) + {1\over 2} \partial_\mu S \partial^\mu S - {M_S^2\over 2} S^2 - {\lambda_S\over 4!} S^4\,,
\end{equation}
below the cut-off $\Lambda$, very similar to Eq.~\eqref{eq:sqcd}, upon expanding the latter. Here, we have also introduced the Lagrangian mass parameter $M_S$ for $S$ for later convenience; the physical (pole) mass of the hidden sector scalar after electroweak symmetry breaking is
\begin{equation}
m_S^2= \frac{M_S^2 + \eta_S v^2/2}{1-\eta_{KS}v^2/\Lambda^2}
\simeq \left(M_S^2 + \frac{\eta_S v^2}{2}\right)\left(1+\frac{\eta_{KS}v^2}{\Lambda^2}\right)
\end{equation}
\end{subequations}
\noindent with the electroweak vacuum expectation value $v\approx 246~\text{GeV}$. $\lambda_S$ refers to the quartic self-coupling of $S$. The effective theory provided by Eq.~\eqref{eq:heftex} matches onto the composite scenario discussed in~\cite{Frigerio:2012uc}.

These interaction terms are reminiscent of a HEFT-like structure in the hidden sector, i.e. the lowest-lying state is described by a singlet that could be coupled in some non-trivial way to a strongly interacting hidden sector below the symmetry-breaking scale. Assuming that the heavy degrees of freedom can communicate with the SM via the portal interaction, a low-energy theory similar to Eq.~\eqref{eq:heftex} can emerge. It is worthwhile mentioning that the operator in $\sim \eta_{KS}$ is unique in the usual sense of EFT categorisation (see also~\cite{Song:2023jqm,Song:2023lxf}). We outline this in appendix~\ref{sec:opcount}. Throughout, we will assume an unbroken ${\mathbb{Z}}_2$ symmetry; $S$ does not obtain a vacuum expectation value $v_S$ at zero temperature.

In the following, we will consider Eq.~\eqref{eq:heftex} as a motivated extension of the standard Higgs portal $\sim \eta_S$. Phenomenologically, this has interesting implications in its own right. Firstly, the portal interactions contribute coherently; $m_S<{m_H}/2$ can be accessed without directly violating existing Higgs signal strength and hidden Higgs decay measurements. 
Secondly, the momentum dependencies lead to a non-decoupling behaviour of the extra scalar that sources additional momentum dependencies in the visible sector through radiative corrections. This, of course, does not occur in renormalisable scenarios and can lead to a modification of Higgs-propagation sensitive observables in addition to characteristic Higgs coupling modifications for $m_S>m_H/2$. Thirdly, the momentum dependence translates to a non-trivial modification of the effective Higgs potential, affecting both the temperature-independent and temperature-dependent parts along with the Daisy corrections in the thermal Higgs potential. These terms are known to be relevant near the critical temperature; therefore, the interactions of Eq.~\eqref{eq:heftex} could have a significant impact on the thermal history of the electroweak scale in the universe, with correlated effects potentially accessible at the LHC.

%%%%%%%%%%%%%%%%%%%
\section{Phenomenological probes at the LHC and beyond}
\label{sec:collider}
%%%%%%%%%%%%%%%%%%%
%%%%%%%%%%%%%%%%%%%
\subsection{Direct sensitivity: Invisible Higgs decay searches}
\label{sec:direct}
%%%%%%%%%%%%%%%%%%%
We first turn to direct sensitivities at the LHC. In contrast to the standard portal (which is of course recovered for $\eta_{KS}=0$), the prompt decay $H\to SS$ is modified for $m_H>2 m_S$
\begin{equation}
\label{eq:decay}
\Gamma(H\to SS) = {1\over 32\pi } \sqrt{1 - {4m_S^2\over m_H^2}} {v^2\over m_H}
\left( \eta_S + \eta_{KS} {2m_S^2 - m_H^2 \over \Lambda^2} 
\right)^2,
\end{equation}
This singles out a new parameter region for which the invisible Higgs decay width, when $m_H> 2 m_S$, is parametrically suppressed for
\begin{equation}
\label{eq:nohid}
\eta_{S} \approx \eta_{KS} {m_H^2 - 2m_S^2 \over \Lambda^2}\,.
\end{equation}
When the standard Higgs portal is open for $m_S\simeq 0$, perturbative (if so tuned) choices of the kinetic portal couplings can remove the Higgs signal constraints from a sizeable invisible decay width for perturbative couplings of the standard portal coupling ($m_H=0.125~\text{TeV},m_S\simeq 0$) 
\begin{equation}
\eta_S \approx 0.016\, \eta_{KS}\,\left(\frac{\Lambda}{\text{TeV}}\right)^{-2}.
\end{equation}
This can be contrasted with unitarity constraints from considering elastic $SH$ scattering using the standard techniques of partial wave projection (see e.g.~\cite{Jacob:1959at,DiLuzio:2016sur}). Perturbative unitarity of $HS\to HS$ up to the cut-off $\Lambda\sim \text{TeV}$ at leading order can be achieved for choices $|\eta_{KS} | /\Lambda^2 \lesssim 7/\text{TeV}^2$ relatively independent of the choice of $m_S$ where it can be expected to leave phenomenological footprints at the LHC (we will consider $m_S<200~\text{GeV}$, see below). An invisible branching ratio can therefore be avoided for $\eta_S\sim 1$ for $\eta_{KS}$ close to the perturbativity limit. 

\begin{figure}
\centering
\includegraphics[width=0.52\textwidth]{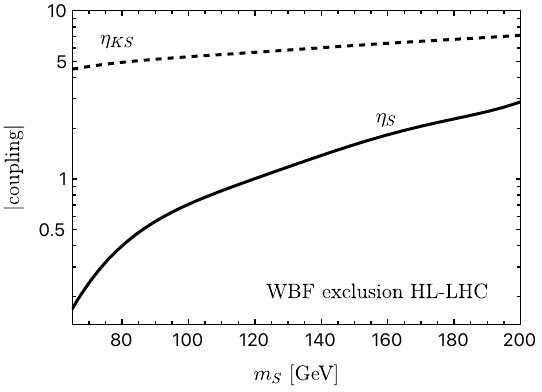}
\caption{\label{fig:wbf}Constraints from a representative invisible Higgs search at ATLAS~\cite{ATLAS:2023tkt} projected to the HL-LHC phase for off-shell production $m_S>m_H/2$. These direct search constraints on $\eta_{KS}$ (assuming $\Lambda=1~\text{TeV}$) start to probe the perturbativity limit of the model for $m_S\gtrsim 200~\text{GeV}$.}
\end{figure}

When viewed as a generic, $\eta_S$-uncorrelated coupling, measurements from invisible Higgs searches impose constraints on the parameter space. Firstly, combined fits to the observed Higgs signal strengths are relevant measurements for this region, see, e.g.~\cite{ATLAS:2016neq}. Furthermore, searches for invisible Higgs decays from jet or $Z$-associated Higgs production~\cite{Davoudiasl:2004aj,Fox:2011pm,Englert:2011us,Englert:2016knz} as well as weak boson fusion (WBF) production~\cite{Eboli:2000ze} provide direct handles on an invisible Higgs decay. WBF typically provides the stronger constraints, and we will focus on this channel in the comparison of sensitivity based on~\cite{ATLAS:2023tkt}, setting an upper limit of 10.7\% on the invisible Higgs branching ratio. This limit is comparable to the constraints from a direct fit to on-shell Higgs signal strengths. However, these direct experimental signatures can also be extended to the $m_H<2m_S$ regime where the extra scalars can be produced through an off-shell Higgs boson~\cite{Ruhdorfer:2019utl,Englert:2020gcp}. Reproducing the limits of Ref.~\cite{ATLAS:2023tkt} and rescaling them with the root of the luminosity, the expected exclusion at the HL-LHC is compared for $\eta_S,\eta_{KS}$ (assuming $\Lambda=1~\text{TeV}$) in Fig.~\ref{fig:wbf}. As can be seen, the expected exclusion for the kinetic portal begins to lose perturbative sensitivity around 200~GeV. In the following, we will therefore consider $m_S\lesssim 200~\text{GeV}$, and we will assume $\Lambda=1~\text{TeV}$ for definiteness. We will revisit the constraints on the invisible decay width below, comparing directly to the indirect constraints that can be obtained in this model at the LHC. The impact of $\lambda_S$ on the visible phenomenology is relatively suppressed compared to the impact of $\eta_S,\eta_{KS}$. In the following, we will therefore set $\lambda_S=0$ without lack of generality, but will return to non-zero $\lambda_S$ when considering phase transitions in Sec.~\ref{sec:cosm}.

%%%%%%%%%%%%%%%%%%%
%%%%%%%%%%%%%%%%%%%

%%%%%%%%%%%%%%%%%%%
\subsection{Indirect sensitivity}
\label{sec:indir}
%%%%%%%%%%%%%%%%%%%
The interactions of Eq.~\eqref{eq:heftex} modify the Higgs boson propagation, creating HEFT-like interactions, built from the SMEFT-invariant $\Phi^\dagger \Phi$. These interactions are typically moved to other sectors when considering the `Warsaw' basis~\cite{Grzadkowski:2010es}; such an approach is not economical in this case. Keeping the effects explicit in the Higgs boson two-point function isolates the phenomenological discovery potential more directly. Interactions will
\begin{enumerate}[label=\arabic*.)]
\item modify visible sector Higgs boson couplings uniformly through wave function renormalisation effects in the on-shell scheme (or equivalently BSM contributions to LSZ factors)~\cite{Englert:2013tya,Craig:2013xia},
\item modify the off-shell behaviour of Higgs propagation,
\item introduce new contributions to multi-Higgs vertex functions, including the effective potential.
\end{enumerate}

There are three relevant processes to consider when examining the second point. Firstly, the tree-level electroweak contribution to four top quark production (see, e.g.,~\cite{Alvarez:2016nrz,Darme:2018dvz,Alvarez:2019uxp,Darme:2021gtt,Blekman:2022jag,Anisha:2023xmh}) receives about a 20\% electroweak correction. The total next-to-leading (NLO) radiative corrections are sizeable~\cite{Frederix:2017wme}, turning four top final states into sensitive BSM tools for discovery, especially when considering Higgs propagator modifications~\cite{Englert:2019zmt}. 
Secondly, $Z$ boson pair production via $gg\to ZZ$ has non-decoupling Higgs contributions due to unitarity~\cite{Kauer:2012hd,Englert:2014aca,Haisch:2022rkm}. Modifications of Higgs boson propagation are therefore a priori relevant. However, the couplings discussed here source oblique Higgs effects that will link vertex and propagator corrections in the broken electroweak phase as observed in~\cite{Englert:2019zmt}. We will find that the qualitative EFT results will generalise to the full propagation of light exotic scalars (see below).
Thirdly, Higgs pair production is sensitive to the interplay of Higgs two- and three-point interactions. The dominant gluon fusion production mode will therefore probe the full array of BSM modifications: momentum-dependent vertex and propagator corrections as well as coupling modifications~\cite{He:2016sqr,Voigt:2017vfz,Englert:2019eyl}. Throughout our analysis, we will assume $\lambda_S=0$; phenomenologically $\lambda_S$ does not play a significant role within expected experimental uncertainties, so that the phenomenology of the processes we analyse is predominantly driven by $\eta_{KS},\eta_S$. We will return to the relevance of $\lambda_S$ in Sec.~\ref{sec:cosm}.

%%%%%%%%%%%%%%%%%%%
\subsubsection{Universal on-shell Higgs coupling modifications}
\label{sec:univ}
%%%%%%%%%%%%%%%%%%%
The interactions considered in this work primarily affect the physical Higgs fluctuations in the vicinity of the vacuum. In this sense, the interactions $\sim \eta_S, \eta_{KS}$ induce a maximally HEFT-like pattern of SMEFT. In particular, the derivative interactions $\sim \eta_{KS}$ induce new operator structures in the Higgs boson two-point function that are parametrised by the HEFT operator $\sim a_{\Box\Box} \Box H \Box H$~\cite{Brivio:2014pfa,Herrero:2021iqt} that appears in the renormalised Higgs two-point vertex function
\begin{equation}
\label{eq:selfenergy}
i\widehat{\Gamma}_{HH}(p^2) = (p^2 - m_H^2) +\Sigma_{HH}^{\text{Loop}}(p^2) +\left(\delta Z_H (p^2-m_H^2) -\delta m_H^2\right)+\frac{2a_{\Box\Box}}{v^2}p^4.
\end{equation}
(The inverse of this, expanded to a given order in perturbation theory, gives the propagator.) It is, therefore, convenient to exploit the $\text{HEFT}\supset\text{SMEFT}$ relation and perform the renormalisation programme within the HEFT (see, e.g.,~\cite{Herrero:2021iqt,Anisha:2024ljc}). This enables us to go beyond a dimension-six truncation of the interactions directly when computing amplitudes. This is relevant because we consider comparably light BSM spectra and the singularity structure sourced by virtual hidden sector scalars, which sources SMEFT operators of higher dimension than six.\footnote{We choose the on-shell renormalisation scheme for fields and masses and the $\overline{\text{MS}}$ scheme for the additional renormalisation constants.} More concretely the finite part of $a_{\Box\Box}$ in the $\overline{\text{MS}}$ scheme is
\begin{equation}
a_{\Box\Box}^{\overline{\text{MS}},\text{fin.}}=
{\eta_{KS}^2\over 64\pi^2}  {v^4\over\Lambda^4} B^{\text{fin}}_0(p^2,m_S^2,m_S^2)\,,
\end{equation}
where $B_0^{\text{fin}}$ is the renormalised Passarino-Veltman~\cite{Passarino:1978jh} two-point function. The HEFT parameter is promoted to a form factor, which will generalise to higher-leg HEFT parameters as a function of the relevant Lorentz-scalar quantities. It is instructive to consider the expansion for heavy scalars in the limit $m^2_S\gg p^2$, which yields
\begin{equation}
a_{\Box\Box}^{\overline{\text{MS}},\text{fin.}} (\mu^2) =
{\eta_{KS}^2\over 64\pi^2}  {v^4\over\Lambda^4}  \log \left({\mu^2\over m_S^2}\right) + {\cal{O}}\left({p^2\over m_S^2}\right),
\end{equation}
with the renormalisation scale $\mu$. In this limit, we recover the expected momentum independence, and this result is also consistent with a renormalisation group flow from matching the SM to the full theory at a scale $m_S$.

We will set the renormalised HEFT coefficients (such as $a_{\Box\Box}$ and additional coefficients relevant for $HH$ production, see below) to zero, imagining there are no additional sources that lead to a finite contribution to $a_{\Box\Box}$ (for extraction strategies see the recent~\cite{Englert:2025xrc}). Finite logarithmic corrections will dynamically source these interactions for process-specific scales $p^2\neq m_H^2$.

%%%%%%%%%%%%%%%%%%%
\begin{figure}[!t]
\centering
\includegraphics[width=0.49\textwidth]{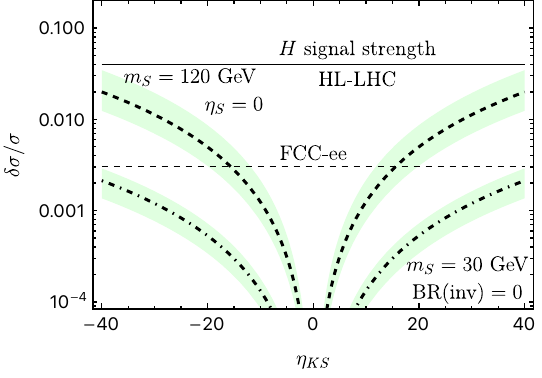}
\caption{\label{fig:hss} Relative Higgs cross section constraints $\delta\sigma / \sigma$ understood as universal Higgs coupling corrections for different colliders and the model discussed in the text.  We show $m_S=30~\text{GeV}$ for parameter choices that remove invisible branching ratio constraints (dot-dashed) through a choice of $\eta_S$. Also shown is $m_S=120~\text{GeV}$ for $\eta_{S}=0$ (dashed). The green bands represent the scale uncertainty, which is obtained from varying the renormalisation scale in $\mu \in [0.5m_H,2m_H]$ for a central choice $\mu=m_H$. Throughout, we choose $\Lambda=1~\text{TeV}$. An optimistic target for the HL-LHC is a 2\% determination of universally rescaled SM Higgs couplings~\cite{Cepeda:2019klc},  which can be improved by a 0.31\% measurement of the Higgs strahlung cross section~\cite{Selvaggi:2025kmd} at a future Higgs machine (here represented by FCC-ee). Other concepts such as the~ILC~\cite{Bambade:2019fyw}, CLIC~\cite{CLICdp:2018cto}, CEPC~\cite{CEPCStudyGroup:2023quu}, or LCF~\cite{LinearColliderVision:2025hlt,LinearCollider:2025lya} can obtain quantitatively similar constraints.}
\end{figure}
%%%%%%%%%%%%%%%%%%%

Owing to the nature of the portal interactions, all ${\cal{O}}(\eta_{KS}^2)$ radiative corrections to SM quantities are given by BSM contributions to SM renormalisation constants; Goldstone propagation and Goldstone vertex corrections cancel identically for physical processes at this order. We can therefore perform the calculation in Feynman gauge $\xi=1$ and afterwards decouple the Goldstone sector in unitary gauge $\xi\to \infty$ for simplicity. This further highlights the physical Higgs properties as the main phenomenological drivers, as expected in HEFT. Carrying out the renormalisation programme (we give details further below), we find universal coupling modifications for fermions $f$ and massive gauge bosons (suppressing the known $\eta_S$ result, e.g.~\cite{He:2016sqr,Voigt:2017vfz,Englert:2019eyl}, for convenience)\footnote{We perform calculations in this work with {\tt{FeynArts}}/{\tt{FormCalc}}/{\tt{LoopTools}}~\cite{vanOldenborgh:1989wn,Mertig:1990an,Hahn:2000kx,Hahn:1998yk,Hahn:2000jm,Shtabovenko:2016sxi,Shtabovenko:2020gxv}.}
\begin{multline}
\kappa^V = \kappa^f = 1+\delta \kappa_H
= \\ 1 + {\eta_{KS}^2 \over 32 \pi^2} {v^2\over \Lambda^4}
\bigg( A^\text{fin}_0(m_S^2) -  (m_H^2 -2 m_S^2)
B^{\text{fin}}_0(m_H^2,m_S^2,m_S^2) + {m_H^2-2 m_S^2 \over 2}
B^{\prime\text{\,fin}}_0(m_H^2,m_S^2,m_S^2) 
\bigg)\,,
\end{multline}
again in terms of the real and renormalised parts of the standard Passarino-Veltman~\cite{Passarino:1978jh} one-loop scalar integrals $A_0,B_0,B'_0$ (and the derivative of the $B_0$ function indicated by the prime). We show the sensitivity from projected single Higgs observations (the relative cross section change $\delta\sigma/\sigma$ due to universal coupling modifications) in Fig.~\ref{fig:hss} for two mass scenarios alongside scale variations in relation to changes of $\mu$. As can be seen, the precision that becomes available at the HL-LHC for single Higgs observables does not improve over the off-shell suppression of direct production in Fig.~\ref{fig:wbf}. When $S$ is light and is characterised such that $H\to SS$ is suppressed through destructive interference, the Higgs coupling modification will fall below the HL-LHC sensitivity threshold. Here, a future $e^+e^-$ Higgs factory could partly regain sensitivity and constrain the kinetic portal for heavier states in parallel. These constraints are, however, close to where we can expect perturbativity to be lost.

%%%%%%%%%%%%%%%%%%%
\begin{figure}[!t]
\centering
{\includegraphics[width=0.49\textwidth]{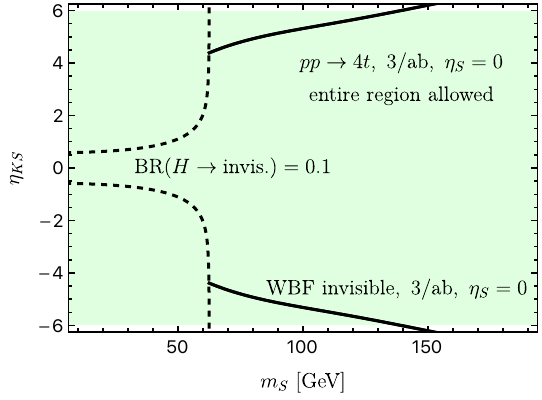}}
\caption{\label{fig:4ta} Constraints from four top quark production on the kinetic Higgs portal with $\Lambda=1~\text{TeV}$. We also show constraints from invisible Higgs constraints imposed by 125 GeV Higgs boson signal strength measurements for $\eta_S=0$ (black, dashed). These are extended by WBF constraints in the off-shell regime (black solid). The production of four top quarks (green shaded region) does not add sensitivity beyond these direct constraints.}
\end{figure}
%%%%%%%%%%%%%%%%%%%

%%%%%%%%%%%%%%%%%%%
\subsubsection{Processes with Higgs off-shell modifications}
%%%%%%%%%%%%%%%%%%%
We now turn to constraining the momentum dependence imparted on the physical Higgs boson's propagation in four top quark, $ZZ$, and $HH$ production. Firstly, we consider $t\bar t t \bar t$, which has recently been observed by ATLAS~\cite{ATLAS:2023ajo} and CMS~\cite{CMS:2023qyl}, surpassing their midterm sensitivity extrapolations. It can be expected that the HL-LHC will further improve its sensitivity to this channel, eventually being able to set relatively tight constraints on new physics~\cite{Belvedere:2024wzg}. Similar to on-shell Higgs production, we renormalise the $t\bar t \to H \to t\bar t $ amplitude. As mentioned above, we can decouple the Goldstone boson contributions so that the off-shell $t\bar t \to t\bar t$ amplitudes can be interfaced with {\tt{MadGraph\_aMC@NLO}}~\cite{Alwall:2014hca}. The result for the HL-LHC\footnote{Throughout this paper, we consider 13.6 TeV collisions for our HL-LHC projections.} is shown in Fig.~\ref{fig:4ta} for the four-top production extrapolation to the HL-LHC of~\cite{Belvedere:2024wzg}. Here we also revisit the combined Higgs signal strength constraints and direct WBF constraints detailed in Sec.~\ref{sec:direct}. The production of four top quarks, unfortunately, does not provide competitive constraints, cf. Fig.~\ref{fig:4ta}.

%%%%%%%%%%%%%%%%%%%
\begin{figure}[!t]
\centering
\includegraphics[width=0.52\textwidth]{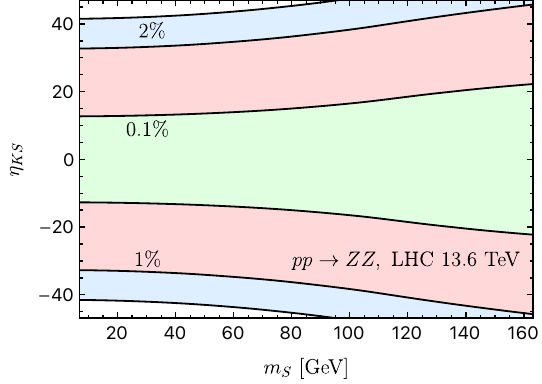}
\caption{\label{fig:zz} Cross section deviations of $gg\to ZZ$ from the SM expectation through its dependence on the modified Higgs propagator, for $\eta_S=0$.}
\end{figure}
%%%%%%%%%%%%%%%%%%%

Next, we consider $ZZ$ production.\footnote{Both $ZZ$ and $HH$ results have been obtained with {\tt{vbfnlo}}~\cite{Arnold:2008rz}.} Similar to the $t \bar t \to t\bar t$ amplitudes the $t\bar t \to ZZ$ amplitude can be renormalised. The universal character of the Higgs portal guarantees a relation between the renormalisation constants\footnote{Electroweak one-loop renormalisation techniques have been reviewed extensively in~\cite{Denner:1991kt,Denner:2019vbn}.} 
\begin{equation}
-\delta Z_H\big|_{\text{div}} = {\delta m_W^2\over m_W^{2} }\bigg|_{\text{div}} = {\delta m_t\over m_t}\bigg|_{\text{div}} 
= - {3\eta_{KS}^2\over 16\pi^2} {m_{S}^2\over \Lambda^2} {v^2\over \Lambda^2}\,,
\end{equation}
(again for $\eta_S=0$ and suppressing the poles in dimensional regularisation $d=4-2\epsilon$) where the renormalisation constants $\delta m_W^2, \delta m_t$ are understood as terms $\sim \Lambda^{-4}$. In particular, a renormalisation of the Weinberg angle is not required. We note the divergent contributions to the Higgs mass renormalisation, as well as the HEFT parameter $a_{\Box\Box}$ for completeness
\begin{align}
\delta m^2_H\big|_{\text{div}} &= {\eta_{KS} \over 16\pi^2} {m_S^4 \over \Lambda^2} \left(
1+ 3\eta_{KS} {2 m_S^2 -m_H^2\over \Lambda^2} {v^2\over m_S^2} \right), \\
\delta a_{\Box\Box}\big|_{\text{div}} &= - {\eta_{KS}^2 \over 64 \pi^2} {v^2\over \Lambda^4}\,.
\end{align}
As for $t\bar t \to t\bar t$ mentioned above, entering these counterterms in the one-loop renormalised amplitude, we obtain a UV-finite result. The $t\bar t \to ZZ$ amplitude then extends to the $gg\to H\to ZZ$ amplitude. The $gg\to ZZ$ amplitude through fermionic box contributions remains SM-like. This way, the $gg \to ZZ$~\cite{Kauer:2012hd,Campbell:2013una,Englert:2014aca} can be generalised to the portal scenario including the virtual $S$ contributions via the $t\bar t\to H \to ZZ$ subamplitude. As the Higgs contribution is relatively small and given the relation of broken and unbroken phase detailed in~\cite{Englert:2019zmt}, the overall corrections are minor, and not large enough to set competitive constraints, see Fig.~\ref{fig:zz}. 

Finally, we turn to Higgs pair production. The renormalisation programme has been described in detail in~\cite{Anisha:2024ljc}. We treat the tadpole contributions in the parameter-renormalised scheme~\cite{Denner:2019vbn}, which requires their inclusion in the renormalisation of the Higgs three-point vertex function (they also contribute to the Goldstone 2-point function). The momentum dependence sourced by the virtual $S$ contributions will renormalise a range of chiral dimension-two and four operators on the HEFT side~\cite{Buchalla:2017jlu}. Without repeating details here, this requires the renormalisation of the trilinear Higgs coupling $\kappa_3$ as well as the HEFT parameters $a_{dd\Box}, a_{Hdd}, a_{H\Box\Box}$ in the basis of~\cite{Herrero:2021iqt}. Again, we choose these couplings to vanish at a given renormalisation scale (for measurement strategies of this input data, we refer the reader to \cite{Englert:2025xrc} once more).

%%%%%%%%%%%%%%%%%%%
\begin{figure}[!t]
\centering
\includegraphics[width=0.49\textwidth]{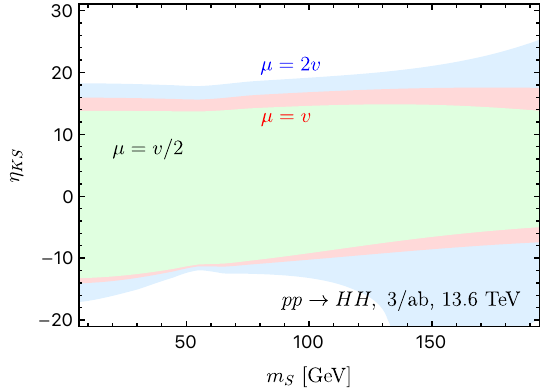}
\caption{\label{fig:hh} Higgs pair production constraints from ATLAS and CMS projections~\cite{CMS:2025hfp} as described in the text. Also shown is the impact of a scale variation $\mu\in [0.5,2]v$ for a central scale choice $\mu=v$. We choose $\eta_S=0$ for illustration.}
\end{figure}
%%%%%%%%%%%%%%%%%%%

In terms of expected limits, ATLAS and CMS have very recently~\cite{CMS:2025hfp} updated their HL-LHC projections across a range of motivated double Higgs final states, setting limits on modifications of the Higgs trilinear couplings within $[-26\%,+29\%]$ at 68\% confidence level. This interval can be mapped onto a cross section constraint using the cross section interpolation of the LHC Higgs Working Group and~\cite{Heinrich:2022idm, Bagnaschi:2023rbx}, and we interpret this cross section constraint $-25\% \lesssim 1-\sigma/\sigma_{\text{SM}} \lesssim 23\%$ onto the parameter space of the model considered here. The result is shown in Fig.~\ref{fig:hh}. In comparison to the other channels discussed so far, Higgs pair production does not provide more competitive constraints and shows a larger vulnerability to scale uncertainties. This is due to the increased relevance of logarithms related to the renormalisation of the multi-Higgs vertex functions (see, e.g.,~\cite{Herrero:2021iqt, Anisha:2024ljc}).
%%%%%%%%%%%%%%%%%%%
\section{Relevance for early universe physics}
\label{sec:cosm}
%%%%%%%%%%%%%%%%%%%
\subsection{The electroweak phase transition}
\label{sec:cosm1}
%%%%%%%%%%%%%%%%%%%
The cosmological relevance of the Higgs portal in its standard form $\sim \eta_S$ has been studied extensively in the literature~\cite{Espinosa:1993bs,Profumo:2007wc,Curtin:2014jma,Carena:2019une,Ramsey-Musolf:2024ykk,Niemi:2024axp}, in particular with regard to its ability to trigger a first-order phase transition in the early universe. These analyses showed that for a strong-first order phase transition (SFOEWPT) in the vanilla portal scenario, a sizeable $\eta_S$ of ${\cal{O}}(1)$ and $\lambda_S>0$ is necessary. This directly pits the SFOEWPT against invisible decay widths (and correlated visible channel signal strength constraints) for $m_S<m_H/2$ as well as universal loop-induced coupling modifications for heavier portal scalars, typically leading to large tension. In both regions, the additional freedom of $\eta_{KS}$, however, can be exploited to relax experimental constraints.

To quantify the phenomenological outcome, we employ a comprehensive scan over the standard portal scenario $\sim\eta_S\in [-1,1]$ ($\eta_{KS}=0$), $m_S \in [0,200]\,\text{GeV}$, $\lambda_S \in [0,6]$ using \texttt{BSMPTv3}~\cite{Basler:2024aaf} to identify a viable SFOEWPT region. We then reevaluate the phenomenology in the light of the $\eta_{KS}$ discussion~above to see if consistent $\eta_{KS}$ choices are possible to remove the tension with other measurements. We will return to the possibility of driving the phase transition through $\eta_{KS}$ further below. For definiteness, we will understand the `strength' of the phase transition as
\begin{equation}
\xi_{p} = \frac{v(T_{p})}{T_{p}}\,,
\end{equation}
where $v(T_{p})$ denotes the Higgs vacuum expectation value at the percolation temperature $T_{p}$. A strong first-order phase transition requires $\xi_{p} > 1$ (additional details can be found in~\cite{Basler:2024aaf}). Depending on the parameter choices, we find first-order transitions with one or two steps between phases with typically non-zero $v_S$, while $v_S = 0$ at zero temperature. The strength $\xi_p$ is evaluated for the step in which the Higgs vacuum $v$ changes from zero to a non-zero value at the corresponding percolation temperature. For parameter choices involving dimension-four couplings, we find viable points exhibiting a strong first-order phase transition ($1< \xi_p <2$) for singlet masses up to $m_{S}= 160~\text{GeV}$.

We start with the parameter region $m_S<m_H/2$. We find in our scan that 
parameter choices for $\eta_{KS}$ that suppress invisible Higgs decays violate the unitarity constraints for our benchmark choice of $\Lambda=1~\text{TeV}$. Hence, a viable SFOEWPT due to $\eta_S$ cannot be compensated with perturbative $\eta_{KS}$ choices. Lower scales are imaginable, but these would lead inadvertently to strong coupling effects dialled into the visible sector as well.

Turning to $m_S > m_H/2$ in our scan, a successful SFOEWPT driven by $\eta_S$ leads to universal Higgs coupling modifications at ${\cal{O}}(\eta_S^2)$ that are large enough to constrain this parameter region at present already. Approaching the HL-LHC phase, the entire parameter region probed in our scan can be ruled out. Again, cancellations at one-loop are possible given the parametric freedom of $\eta_{KS}$. And for our $\eta_S\neq 0$ scan points, these are within the perturbative range, we find $|\eta_{KS}|/\Lambda^2 \lesssim 6/\text{TeV}^2$ just below the perturbativity limit quoted above. Through an appropriate choice of $\eta_{KS}$, cancelling the impact of $\eta_S\neq0$, the single Higgs observables will not be sensitive at the one-loop level, highlighting double Higgs production as the remaining potentially sensitive collider probe of this parameter region. For such choices, the double Higgs production cross section modifications are, however, below 1\%, compared to the SM expectation; the bulk of the (ad hoc) single Higgs phenomenology cancellations carry over to multi-Higgs final states. These cross section modifications are too small to be observable at the LHC. Therefore, within the scope of our analysis, we are forced to conclude that whilst a $\eta_S$-driven SFOEWPT can be obtained in the kinetic portal scenario, this seems only possible through a delicate balance of coupling choices close to the strong coupling limit that leave no additional collider sensitivity otherwise --- a theoretically unappealing avenue.

Next, we investigate how $\eta_{KS}$ reshapes phase transitions and whether there is a possibility to realise an SFOEWPT through {\emph{a combination}} of $\eta_{KS},\eta_S$ effects. Including the non-trivial kinematic dependence $\eta_{KS}$, the one-loop effective potential\footnote{We refer the reader to Ref.~\cite{Balui:2025kat} for recent and new insights into the effective potential's gauge-(in)dependence.} arising from $S$ exchange is
\begin{equation}
V_S^{(1)} (\phi_C) = i \sum_{n=1}^\infty \int {\hbox{d}^4 q \over (2\pi)^4} {1\over 2n}
\left[ {(\eta_S + 2\eta_{KS} \,q^2/\Lambda^2)  \over q^2 - M_S^2 +i\varepsilon }{\,\phi_C^2\over 2}\right]^n\,,
\end{equation}
upon summing all 1-particle irreducible $S$ insertions at one-loop (where $\phi_{C}$ is the background Higgs field). The calculation can be tackled with standard techniques, e.g.~\cite{Quiros:1999jp}, which yields a volume effect
\begin{equation}
\label{eq:cwn}
V^{(1)}_S (\phi_C) = \alpha^{-4}(\phi_C,\eta_{KS})\, V^{(1)}_S(\phi_C,\eta_{KS}=0)\,,
\end{equation}
with
\begin{equation}
\alpha^2(\phi_C,\eta_{KS})= 1- \eta_{KS}{\phi_C^2 \over \Lambda^2}\,.
\end{equation}
Here, $V^{(1)}_S(\phi_C,\eta_{KS}=0)$ denotes the `standard' portal interaction Coleman-Weinberg potential, which is trivially recovered for $\eta_{KS}=0$. 

The finite temperature $T\neq 0$ contribution can be computed similarly. It is given by
\begin{equation}
\begin{split}
\label{eq:ftemp}
V^{(1)}_{S,T}\left(m_S(\phi_C)\right) = {1\over \alpha^4} {\alpha^4 T^4\over 2\pi^2} J_B\left({m_S^2(\phi_C) \over  \alpha^2T^2} \right)
&=
{1\over \alpha^4} V^{(1)}_{S,\alpha T}(m_S(\phi_C),\eta_{KS}=0) \\
&=V^{(1)}_{S, T}({m_S(\phi_C)}/\alpha,\eta_{KS}=0) \,,
\end{split}
\end{equation}
with the standard bosonic $J_B$ function, see e.g.~\cite{Quiros:1999jp}. Again, this reproduces the standard portal result for vanishing $\eta_{KS}$. 

Effectively, $\eta_{KS}$ leads to a change in inertia that exerts itself as an overall modification of energy densities (as it is equivalent to a scale transformation in momentum space), as well as through an effective change in temperature of the thermal bath. When the effective inertia is small, modes $m_S(\phi_C)$ are easily excited, in a given thermal bath and background field $\phi_C$. The temperature that characterises the same free energy of the plasma appears lower compared to the $\alpha=1$ case. Equivalently, as $m_S^2(\phi_C)$ breaks classical scale invariance, at finite temperature, the potential can also be understood as an increased effective mass for $\alpha<1$ at the same temperature $T$ that particles characterised by $\alpha=1$ feel.

Additionally, there are thermal mass corrections to the SM scalar bosons that need to be included via the Daisy resummation. In the high-temperature limit, these are 
\begin{equation}
\Delta\overline{m}^2 = {\eta_{KS} \over 12} {M_{S}^2 \over \Lambda^2} T^2.
\end{equation}
The scalar $S$ does not receive additional thermal corrections beyond the standard portal interaction $\sim \eta_S$ and its self-coupling that are well-documented in the literature~\cite{Quiros:1999jp}. 

%%%%%%%%%%%%%%%%%%%
\begin{figure}[!t]
	\centering
	\includegraphics[width=\textwidth]{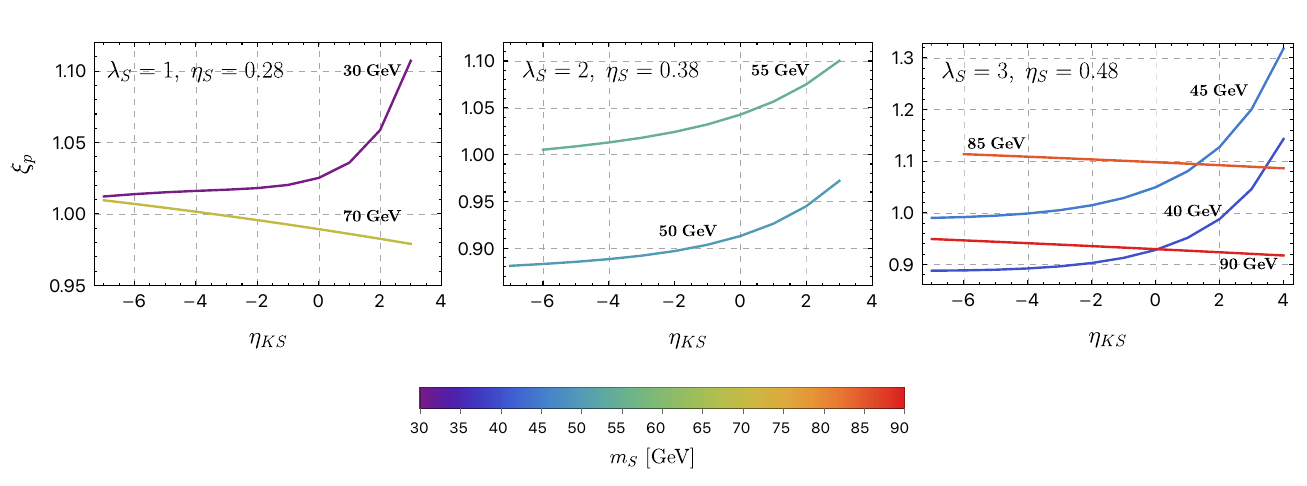}\vspace{-7mm}
	\caption{\label{fig:PT} Response of the phase transition strength $\xi_{p}$ with varying $\eta_{KS}$ for $\Lambda=1~\text{TeV}$. The plots are shown for three chosen benchmark points, with different values of $m_{S}$ indicated by the colorbar. The figure panels show $\xi_{p}$: leftmost, for $m_{S}=30$ and $70$ GeV with $\lambda_S=1,\,\eta_S=0.28$; center, $m_{S}=50$ and $55$ GeV with $\lambda_S=2,\,\eta_S=0.38$; rightmost,  $m_{S}=40, 45, 85, 90$ GeV for $\lambda_S=3,\,\eta_S=0.48$.}
\end{figure}
%%%%%%%%%%%%%%%%%%%

To explore the parameter space relevant for SFOEWPTs numerically, the above-mentioned modifications to the effective potential are implemented in \texttt{BSMPTv3}. We again deploy our previously mentioned \texttt{BSMPTv3} scan without imposing collider constraints at this stage. The parameter points with $\xi_{p}\sim 1$ (allowing for a $10\%$ variation) are then selected to study the impact of an EFT portal coupling, which is varied within its perturbative limit, $|\eta_{KS}| \leq 7$, for $\Lambda=1\,\text{TeV}$.\footnote{If $\eta_{KS}$ is to be consistent as a perturbation, we have to rely on portal scenarios at dimension-four level that show already a relatively strong phase transition, so that a perturbative approach remains valid.}

For lighter singlet masses $m_{S}\leq 55~\text{GeV}$, we observe that the phase transition strength increases with increasing $\eta_{KS}$, whereas for $m_{S}>55~\text{GeV}$ (up to maximum $m_{S}=155~\text{GeV}$), it decreases. To illustrate the impact of $\eta_{KS}$, we selected three benchmarks with fixed $(\lambda_S, \eta_S)$ and different singlet masses. The corresponding response is shown in Fig.~\ref{fig:PT}. We note that the light mass choices, for which $\xi_p>1$ is achieved via $\eta_{KS}$, are not compatible with constraints from the 125 GeV Higgs signal strength measurements due to a large invisible Higgs decay width $H\to SS$. These parameter choices are also in tension with direct detection experiments (see below). Whilst driving an SFOEWPT does not seem possible within the constraints of the effective model analysed here, we can expect that a realistic theory of a strongly-interacting hidden sector contains more degrees of freedom that can drastically change the conclusions presented here for a single interpolating field.

%%%%%%%%%%%%%%%%%%%
\subsection{Dark matter relic abundance and direct detection}
\label{sec:cosm2}
%%%%%%%%%%%%%%%%%%%
Understanding the $\mathbb{Z}_2$-odd scalar as a minimal solution to the WIMP miracle is experimentally challenged. The generic finding of the standard singlet scenario is that the concordance of dark matter relic abundance $\Omega_{\text{DM}}h^2\simeq 0.12$ and direct detection exclusion leads to a substantial tension. If the hidden sector is less minimal and contains additional fields, this tension can be reduced. Nonetheless, to gauge the compatibility of the discussed scenario with these data, we will assume here that $S$ is indeed stable and the only relevant state for direct detection and relic abundance computations. Similar to extending the parameter space via the kinetic interactions $\sim\eta_{KS}$ into the regime $m_S \leq m_H/2$, we can then also revisit the implications for astrophysics. We employ {\tt{micrOMEGAs}}~\cite{Alguero:2023zol} to identify regions of the $(m_S,\eta_S,\eta_{KS})$ parameter space where the correct DM relic abundance is reproduced whilst no direct detection constraints can be obtained. In this region $m_S\lesssim m_H/2$, we find that a choice
\begin{equation}
m_S= 55~\text{GeV},~\eta_{KS}=-0.3,~\eta_S=0.003,~(\text{BR(inv)}=1.9\%)\,,
\end{equation}
is consistent with experimental observations and can approximate the astrophysical data within 10\%. Viewed against our previous discussion of Sec.~\ref{sec:cosm1}, it is clear that this region is not compatible with the simultaneous requirement of an SFOEWPT. The viable region in parameter space for $m_S\lesssim m_H/2$ is small and well-represented by this point. The coupling deviations for this parameter point will also be observable at a future Higgs factory~\cite{Selvaggi:2025kmd}; the invisible branching ratio exceeds the (representative) 0.31\% accuracy obtainable for Higgs production at a lepton collider.\footnote{As part of the renormalisation programme detailed in the previous section, we have checked that the RGE flow does not significantly change the correlation between measurements at the involved different energy scales.} Turning to regions $2m_S> m_H$, around the threshold region, much bigger $\eta_{KS}$ are allowed, reaching $|\eta_{KS}|\simeq 4.5$ around 70~GeV; relic abundance and direct detection results are reproduced by $\eta_{S}\simeq 0.04$. The associated coupling deviations are, however, below the indirect sensitivity that can be established at present and the next generation of precision colliders, such as FCC-ee.

The phenomenological possibility of simultaneously addressing these constraints with the interactions considered here is not new. It is the very motivation behind the aforementioned Composite Dark Matter theories. The astrophysical implications of these models have been discussed in great detail elsewhere~\cite{Frigerio:2012uc,Marzocca:2014msa,Fonseca:2015gva,Bruggisser:2016ixa}. The qualitative features revealed in our scan are explained in these models as their shift symmetry suppresses elastic scattering of on-shell $S$ particles and, therefore, direct detection constraints are avoided~\cite{Frigerio:2012uc}. The shift symmetry is hidden in our parametrisation of Eq.~\eqref{eq:heftex} but reproduced in appendix~\ref{sec:opcount}. Crucially, it identifies the regions quoted in our scan as parameter choices that are motivated by composite theories. 

%%%%%%%%%%%%%%%%%%%
\section{Summary and conclusions}
\label{sec:conc}
%%%%%%%%%%%%%%%%%%%
Non-minimal Higgs sector extensions arise in a multitude of BSM theories. In this work, we have focused on strongly interacting hidden sectors that give rise to non-standard effective modifications of the so-called Higgs portal. The leading effects of such interactions can be motivated from chiral perturbation theory and its large $N$ generalisation via AdS/CFT, and they are kinetic in nature, introducing unique non-standard momentum dependencies at the dimension-six level. Concrete examples of this have been widely studied in the literature as Composite Scalar Dark Matter, where parameter choices in our bottom-up effective field theory treatment emerge through symmetry. In contrast, we have dissected visible, loop-induced signatures of hidden-sector momentum-dependencies via Higgs Effective Field Theory-inspired methods in Sec.~\ref {sec:collider}. 

More concretely, non-standard momentum dependencies of portal interactions manifest themselves predominantly through modifications of the physical Higgs boson. In this sense, the phenomenology is well-captured in a HEFT approach, albeit communicated to the SM sector in a SMEFT-like fashion. We find that the effects of the kinetic portal extension can be probed via universal Higgs coupling modifications. Due to its indirect sensitivity, this information is insufficient to reveal the momentum-dependent nature of the extension in the presence of standard, renormalisable couplings. The non-decoupling nature of these interactions (below the hidden sector's intrinsic mass scale, here taken to be ${\cal{O}}(\text{TeV})$), however, highlights other phenomenological arenas for potential sensitivity, as they can leave radiative imprints on visible-sector observables. We demonstrate that some sensitivity can be found in Higgs pair production (yet close to the unitarity constraint). $Z$ pair and four top production are largely insensitive to the presence of these states.

Such non-trivial momentum dependencies can impact the physics of the early Universe.\footnote{They will also affect more broadly astrophysical observations related to strongly interacting dark sectors such as the cusp vs. core anomaly~\cite{deBlok:2009sp}, the `too big to fail' problem~\cite{Boylan-Kolchin:2011qkt} or the missing-satellites problem~\cite{Klypin:1999uc}, e.g. by modifying galaxy core formation and feedback.} Especially when the exotic hidden sector scalar is light, the electroweak phase transition can receive significant modifications compared to the SM-expected crossover. Thus, for lighter hidden sector scalars, the strength of the phase transition can be modified to first-order through the $\eta_{KS}$ coupling. This region is tensioned by Higgs signal strength measurements and invisible Higgs decay searches. For heavier $S$, $\eta_{KS}$ has a reduced relevance for driving an SFOEWPT.

Furthermore, the higher-dimensional interactions enable the Higgs portal's ability to act as a viable dark matter candidate, satisfying relic abundance and direct detection constraints, in line with the Composite Scalar Dark Matter paradigm. The relevant parameter region can be explored at a precision lepton collider such as the widely-discussed FCC-ee, through modifications of the Higgs width and a small invisible branching ratio, that is beyond the sensitivity of the LHC.

%%%%%%%%%%%%%%%%%%%
\subsubsection*{Acknowledgements}
%%%%%%%%%%%%%%%%%%%
We thank Victor Maura Breick, Max Detering, Suraj Prakash and Tevong You for helpful discussions on matters related to this work. CE is particularly indebted to Max Detering for discovering a numerical error in parts of the results presented in an earlier version of this work.
We thank Christoph Borschensky for helpful discussions and for providing a {\tt BSMPT} model implementation for the real scalar singlet extension on which our model implementation is based.
We also thank the journal referee for their comments and suggestions which helped improve this work.
A. and M.M. acknowledge support by the Deutsche Forschungsgemeinschaft (DFG, German Research Foundation) under grant 396021762 - TRR 257.
The work of L.B. is supported by the Swiss National Science Foundation (SNSF).
C.E. is supported by the STFC under grant ST/X000605/1, and by the Leverhulme Trust under Research Fellowship RF-2024-300$\backslash$9. C.E. is further supported by the Institute for Particle Physics Phenomenology Associateship Scheme.
%
%%%%%%%%%%%%%%%%%%%
\appendix
%%%%%%%%%%%%%%%%%%%
\section{Counting derivative portal interactions at leading order}
\label{sec:opcount}
%%%%%%%%%%%%%%%%%%%
For derivative operators of class $\phi^4 D^2$ (with $\phi=\Phi,S$) involving a ${\mathbb{Z}}_2$ symmetric singlet, an additional invariant structure $\Phi^2 S^2 D^2 $ (see Refs.~\cite{Song:2023jqm,Song:2023lxf}) is possible using the Hilbert series method, in addition to the SMEFT $ \Phi^4 D^2$ operators. The following gauge-invariant structures with two derivatives are possible in combination with the gauge-singlet Higgs operator $\Phi^\dagger \Phi$
\begin{gather}
\Phi^\dagger \Phi \partial_{\mu} S \partial^{\mu}S\,, \label{eq:1}\\
\partial_\mu(\Phi^\dagger \Phi) S \partial^{\mu}S\,,\label{eq:2}\\
\Box(\Phi^\dagger \Phi) S^2\,, \label{eq:3}\\
\Phi^\dagger \Phi  S \Box S\,.\label{eq:4}
\end{gather}
The last of these, Eq.~\eqref{eq:4} $\sim \Box S$, is linked to the equations of motion (EOM) of $S$ and in systematic expansion in the EFT scale, these interactions are removed~\cite{Grzadkowski:2010es} as these reduce to  $\phi^4$ (dim-4 portal $\Phi^\dagger \Phi S^2$) and $\phi^6$ structures.\footnote{As EOMs are not equivalent to redundant field redefinitions, the theories obtained this way are not strictly identical~\cite{Criado:2018sdb}. A complete renormalisation programme of off-shell Green's functions at a given order requires additional care. This has been transparently demonstrated in Ref.~\cite{Herrero:2022krh}.} We can rewrite Eq.~\eqref{eq:3} using integration by parts (IBP), e.g.
\begin{equation}
\label{eq:strip}
\Box (\Phi^\dagger \Phi) S^2 
\to \partial_\mu (\Phi^\dagger \Phi) \partial^\mu (S^2)
\to \partial_\mu (\Phi^\dagger \Phi) S \partial^\mu S\,.
\end{equation}
Similarly, starting from $\partial_\mu (\Phi^\dagger \Phi S \partial^\mu S)$, we have  
\begin{equation}
	\partial_\mu (\Phi^\dagger \Phi) S\partial^{\mu}S
	\to (\Phi^\dagger \Phi) S\Box S
	+ (\Phi^\dagger \Phi) \partial_\mu S \partial^{\mu}S\,,
\end{equation}
and thus we can also omit Eq.~\eqref{eq:2}. This leaves Eq.~\eqref{eq:1} as the non-redundant structure for the lowest-order momentum deformation of the Higgs portal. Therefore, for the results presented in this work, it is sufficient to consider the interaction~\eqref{eq:1}.

It is worthwhile highlighting that the middle operator in Eq.~\eqref{eq:strip} is the canonical basis choice in composite theories endowed with shift symmetries. In the HEFT-inspired basis choice used in this present work, this operator translates via IBP and EOM into
\begin{equation}
{1\over 2 \Lambda^2} \partial_\mu (\Phi^\dagger \Phi) \partial^\mu (S^2) 
\to  -{1\over \Lambda^2} \Phi^\dagger \Phi \, \partial_\mu S \partial^\mu S +   { M_S^2\over \Lambda^2}\,\Phi^\dagger \Phi \, S^2\,.
\end{equation}
In our parametrisation of Eq.~\eqref{eq:heftex}, this operator is recovered for
\begin{equation}
{\eta_{S}\over 2}  = -{\eta_{KS}} {M_S^2\over \Lambda^2}\,,
\end{equation}
which also implies $m_S=M_S$, i.e. the bracket in Eq.~\eqref{eq:heftex} vanishes for on-shell $S$, which also removes direct detection constraints.
%%%%%%%%%%%%%%%%%%%
%\bibliographystyle{JHEP}
%\bibliography{references}
\bibliography{draft}
%%%%%%%%%%%%%%%%%%%

\end{document}